\documentclass[12pt]{article}
\usepackage[top=26truemm,bottom=26truemm,left=20truemm,right=20truemm]{geometry}
\usepackage{hyperref}
\usepackage[pdftex]{graphicx}
\usepackage{amssymb}
\usepackage{braket}
\usepackage{amsmath}
\newtheorem{defi}{Definition}[section]

\newcommand\eq[1]{\begin{equation}#1\end{equation}}

\newcommand\al{\alpha}
\newcommand\be{\beta}

\newcommand\cb{c_{\beta}}
\newcommand\cg{c_{\gamma}}
\newcommand\ga{\gamma}

\newcommand\om{\omega}

\makeatletter

\newcommand\eqn[1]{\begin{eqnarray}#1\end{eqnarray}}
\@addtoreset{equation}{section}
\makeatother

\begin{document}

\title{A New Way to Derive the Taub-NUT Metric \\with Positive Cosmological Constant\thanks{Alberta-Thy-02-16, arXiv:mmyy.nnnnn[gr-qc]} }
\author{Kento Osuga\thanks{Email address: osuga@ualberta.ca}  \;and\; Don N. Page\thanks{Email address: profdonpage@gmail.com}\\[.5cm] Theoretical Physics Institute \\ Department of Physics\\ 4-181 CCIS\\ University of Alberta \\ Edmonton, Alberta T6G 2E1\\ Canada}

\date{March 17, 2016}
\maketitle

\begin{abstract}
We investigate a biaxial Bianchi IX model with positive cosmological constant, which is sometimes called the $\Lambda$-Taub-NUT spacetime, whose exact solution is well known. The minisuperspace of biaxial Bianchi IX models admits two non-trivial Killing tensors that play an important role for deriving the Taub-NUT metric. We also give a brief discussion about the asymptotic behaviour of Bianchi IX models.

\end{abstract}

\section{Introduction}
The metric of a Bianchi IX spacetime is given as
\eq{ds^2=-dt^2+a^2(t)\om_1^2+b^2(t)\om_2^2+c^2(t)\om_3^2,\label{Bianchi metric}}
where ($t,a,b,c$) have dimensions of time or length, since we are setting the speed of light $c$ (not the $c(t)$ above) equal to unity, as well as Newton's gravitational constant $G$. $\{\om_k\}$ is a set of ${\mathbb S}^3$-invariant one-forms,
\eqn{\om_1&=&\cos\psi\,d\theta+\sin\psi\sin\theta\,d\phi,\nonumber\\
\om_2&=&\sin\psi\, d\theta-\cos\psi\sin\theta\,d\phi,\label{basis}\\
\om_3&=&d\psi+\cos\theta\,d\phi,\nonumber}
obeying
\eq{d\om_1=\om_2\wedge\om_3\hspace{10mm}{\rm et \;cyc}.,}
where $0\leq\theta\leq\pi$, $0\leq\phi<2\pi$, $0\leq\psi<4\pi$. The three principal circumferences of the distorted ${\mathbb S}^3$ are then $(4\pi a, 4\pi b, 4\pi c)$. On the other hand, the $\Lambda$-Taub-NUT spacetimes are originally introduced by \cite{Taub original} as spacetimes whose spatial topology is a biaxial ${\mathbb S}^3$ and which satisfy the Einstein equations with positive cosmological constant,
\eq{R_{\mu\nu}=\Lambda g_{\mu\nu}.\label{Einstein equations}}
The form of the solution depends on the choice of time coordinate; for example it is given by \cite{Taub original 2,Taub consistent} with an arbitrary constant $C_0$ and arbitrary positive constant $D_0$ as
\eq{ds^2=\frac{3D_0}{\Lambda}\left(-\frac{d{\tau}^2}{f({\tau})}+\frac{f({\tau})}{4}\,\om^2_3+\frac{{\tau}^2+1}{4}\,(\om^2_1+\om^2_2)\right),\label{Taub known}}
\eq{f(\tau)=\frac{D_0\tau^4+2(3D_0-2)\tau^2+C_0\tau+4-3D_0}{1+\tau^2}.}
Thus the $\Lambda$-Taub-NUT spacetime is none other than a biaxial Bianchi IX spacetime with positive cosmological constant. In this paper we present a new derivation of (\ref{Taub known}) by considering the minisuperspace defined below.


\section{Minisuperspace}
The orthonormal components of the Ricci tensor of a Bianchi IX spacetime are given by \cite{Ricci} as
\eq{R_{0}^0=\frac{\ddot{a}}{a}+\frac{\ddot{b}}{b}+\frac{\ddot{c}}{c},} 
\eq{R_{1}^1=  \frac{\ddot{a}}{a} +\frac{\dot{a}}{a} \left( \frac{\dot{b}}{b}+\frac{\dot{c}}{c} \right) + \frac{a^4-(b^2-c^2)^2}{2a^2b^2c^2}.}
Here an overdot is a derivative with respect to the proper time $t$ in the metric \eqref{Bianchi metric}. $R_{2}^2$ and $R_{3}^3$ are just permutations of $R_{1}^1$, and off-diagonal elements of the Ricci tensor are zero. Let us write the dimensional variables $(t,a,b,c)$ in terms of dimensionless Misner variables $(\zeta,\al,\be_a,\be_b,\be_c,\be,\ga)$ \cite{Misner:1969ae} as
\eq{t=\sqrt{\frac{3}{\Lambda}}\zeta, \;\;a=\sqrt{\frac{3}{\Lambda}}e^{\alpha+\beta_a}, \;\;b=\sqrt{\frac{3}{\Lambda}}e^{\alpha+\beta_b},\;\; c=\sqrt{\frac{3}{\Lambda}}e^{\alpha+\beta_c},}
\eq{\beta_a=\beta+\sqrt{3}\gamma,\;\; \beta_b=\beta-\sqrt{3}\gamma,\;\; \beta_c=-2\beta,}
where $\al$ tells how spatially large the model is, since the ${\mathbb S}^3$-volume is $16\pi^2(3/\Lambda)^{\frac{3}{2}}e^{3\al}$, while $\be$ and $\ga$ describe how distorted ${\mathbb S}^3$ is. 

The scalar curvature of the distorted ${\mathbb S}^3$ at one time is
\eq{^{(3)}R=\frac{(a+b+c)(-a+b+c)(a-b+c)(a+b-c)}{2a^2b^2c^2}.}
It is interesting to note from Heron's formula that when the numerator is positive, it is the square of the area of a triangle of edge lengths $(2a,2b,2c)$, which are the principal circumferences divided by $2\pi$. Then $^{(3)}R$ is 8 times the square of the area of a triangle of edge lengths the principal circumferences $(4\pi a, 4\pi b, 4\pi c)$, divided by the square of the ${\mathbb S}^3$-volume that is $16\pi^2abc$. When the numerator is negative, so $^{(3)}R<0$, the edge lengths do not obey the triangle inequality. Furthermore, when the triangle exists and is acute, all three of the ${\mathbb S}^3$ Ricci tensor eigenvalues, $^{(3)}R^1_1$, $^{(3)}R^2_2$, and $^{(3)}R^3_3$, are positive, but when the triangle is obtuse or does not exist, two of the Ricci tensor eigenvalues are negative.

Multiplying the ${\mathbb S}^3$ scalar curvature by a quantity proportional to the two-thirds power of the ${\mathbb S}^3$ volume gives the dimensionless quantity
\eq{V=\frac{1}{6}(abc)^{\frac{2}{3}} \,^{(3)}R=\frac{1}{2\Lambda}e^{2\al} \,^{(3)}R=\frac{1}{12}\biggl(4e^{-2\beta}\cosh2\sqrt{3}\gamma-4e^{4\beta}\sinh^22\sqrt{3}\gamma-e^{-8\beta}\biggr).}
Now letting an overdot denotes a derivative with respect to the dimensionless time coordinate $\zeta$, the Einstein equations (\ref{Einstein equations}) give three dimensionless 2nd-order equations
\eqn{\ddot{\alpha}&=&3-3\dot{\al}^2-2Ve^{-2\al},\label{eqa}\\
\ddot{\beta}&=&-3\dot{\alpha}\dot{\beta}+\frac{1}{2}\frac{\partial V}{\partial \beta}e^{-2\alpha},\label{eqb}\\
\ddot{\gamma}&=&-3\dot{\alpha}\dot{\gamma}+\frac{1}{2}\frac{\partial V}{\partial \gamma}e^{-2\alpha},\label{eqc}}
and one dimensionless 1st-order constraint equation,
\eq{\dot{\alpha}^2-\dot{\beta}^2-\dot{\gamma}^2=1-V e^{-2\alpha}.\label{constraint}}
Note that by combining \eqref{constraint}, its time derivative, and any two of \eqref{eqa}-\eqref{eqc}, one can derive the remaining 2nd-order equation, so only the 1st-order constraint \eqref{constraint} and any two of the three 2nd-order equations are independent.

Note also that if we choose $\ga=\dot{\ga}=0$ as an initial condition, then $\partial V/\partial\ga=0$ and $\ddot{\ga}=0$, so $\ga$ remains zero for all time, which is just a biaxial model.

One notices that (\ref{eqa})-(\ref{constraint}) are reproduced by the following action:
\eq{S=\frac{1}{2}\int d\tau\left(N^{-1}e^{3\alpha}(-\dot{\alpha}^2+\dot{\beta}^2+\dot{\gamma}^2)-N\left(e^{3\al}-e^{\al}V\right)\right),\label{ADM}}
where now the dot denotes the derivative with respect to $\tau$, and $N$ is a Lagrange multiplier. The relation between $\zeta$ and $\tau$ is
\eq{\frac{d}{d\zeta}=\frac{1}{N}\frac{d}{d\tau}.}
If we define
\eq{\eta=N\left(e^{3\al}-Ve^{\al}\right),}
then (\ref{ADM}) becomes
\eq{S=-\frac{1}{2}\int d\tau\left(\eta^{-1}\left(e^{6\al}-Ve^{4\al}\right)(-\dot{\al}^2+\dot{\be}^2+\dot{\ga}^2)-\eta\right).\label{superminispace}}
This is a relativistic point-particle action in three dimensions ($\al,\be,\ga$) with mass $m=1$ and the minisuperspace metric
\eq{ds^2=\left(e^{6\al}-Ve^{4\al}\right)(-d\al^2+d\be^3+d\ga^2).}
This three-dimensional (or two-dimensional for biaxial case) curved space obtained from the four-dimensional Bianchi IX space is an example of a metric on minisuperspace whose geodesics give solutions of Einstein equations \cite{DeWitt:1969uf}. Therefore, time evolution of a Bianchi IX space with $\Lambda$ is equivalent to particle motion along a geodesic curve in this minisuperspace. A more rigorous way to obtain (\ref{superminispace}) is shown for example in \cite{ADM ref}.


\section{Killing Tensors}

We now investigate geometrical properties of the minisuperspace associated with the biaxial Bianchi IX model with $\gamma = 0$, so $a = b = (3/\Lambda)^{1/2}
e^{\alpha + \beta}$ and $c = (3/\Lambda)^{1/2} e^{\alpha - 2\beta}$. Let us first define null coordinates ($u,v$) as
\eq{u=\al-\be+\frac{1}{2}\ln3-\frac{2}{3}\ln2,\hspace{5mm}v=\al+\be+\frac{1}{2}\ln3.}
Then the minisuperspace metric, the nonzero terms of the Levi-Civita connection, and the Ricci scalar are respectively
\eq{ds^2=-\frac{4}{27}\,U(u,v)\,dudv\hspace{3mm}{\rm with }\hspace{3mm}U(u,v)=e^{3u+3v}-e^{3u+v}+e^{6u-2v},\label{metric}}
\eq{\Gamma^u_{uu}=U^{-1}\partial_uU,\hspace{5mm}\Gamma^v_{vv}=U^{-1}\partial_vU,}
\eq{R=81\,U^{-3}e^{9u}(3e^{-v}-5e^{v}).\label{Ricci scalar}}

\subsection{Killing Vectors}

We first show the non-existence of a Killing vector. If $K$ is a Killing vector, its components satisfy the Killing equations
\eqn{\nabla_uK_{u}=\partial_uK_u-\Gamma^u_{uu}K_u=U\partial_u(U^{-1}K_u)=0,\\\nabla_vK_{v}=\partial_vK_v-\Gamma^v_{vv}K_v=U\partial_v(U^{-1}K_v)=0,\\\nabla_vK_{u}+\nabla_uK_{v}=\partial_vK_u+\partial_uK_v=0.}
The first two give, for arbitrary functions $f(v)$ and $g(u)$,
\eq{K_u=f(v)U,\hspace{10mm}K_v=g(u)U,}
or equivalently
\eq{K^u=-\frac{27}{2}g(u),\hspace{10mm}K^v=-\frac{27}{2}f(v).\label{vec1}}

In two dimensions, the fact that a Killing vector obeys $\nabla_{\mu}K^{\mu}=0$ implies that the dual one-form
\eq{L=\varepsilon_{\mu\nu}K^{\mu}dx^{\nu}=-f(v)Udu+g(u)Udv}
is closed, $dL=0$, so that locally it is exact, $L=dM(u,v)$. Integrating this along a line of constant $v$ gives
\eq{M=-f(v)\int Udu=-f(v)\left(\frac{e^{3u+3v}}{3}-\frac{e^{3u+v}}{3}+\frac{e^{6u-2v}}{6}\right)+C(v),}
whereas integrating this along a line of constant $u$ gives
\eq{M=g(u)\int Udv=g(u)\left(\frac{e^{3u+3v}}{3}-e^{3u+v}-\frac{e^{6u-2v}}{2}\right)+D(u).}
Only for $f(v)=g(u)=0$ do these agree, so there is no nonzero Killing vector.


\subsection{Rank-2 Killing Tensors}
For a rank-2 Killing tensor, the Killing equations are
\eqn{&&\nabla_uK_{uu}=0,\\&&\nabla_vK_{vv}=0,\\&&\nabla_uK_{vv}+2\nabla_vK_{uv}=0,\\&&\nabla_vK_{uu}+2\nabla_uK_{uv}=0.}
Similar to the calculations for a Killing vector, the first two equations give
\eq{K_{uu}=f(v)U^2,\hspace{10mm}K_{vv}=g(u)U^2,}
while the last two indicate
\eq{K_{uv}=A(u,v)U,}
\eqn{\partial_vA=-\frac{1}{2}\frac{dg}{du}U-g\partial_uU,\label{A1}\\\partial_uA=-\frac{1}{2}\frac{df}{dv}U-f\partial_vU.\label{A2}}
If one integrates (\ref{A1}) with respect to $v$, $A$ becomes
\eq{A=a(u)-\frac{1}{2}\frac{dg}{du}\left(\frac{e^{3u+3v}}{3}-e^{3u+v}-\frac{e^{6u-2v}}{2}\right)-g(u)\left(e^{3u+3v}-3e^{3u+v}-3e^{6u-2v}\right),\label{AA1}}
while from (\ref{A2}), we have
\eq{A=b(v)-\frac{1}{2}\frac{df}{dv}\left(\frac{e^{3u+3v}}{3}-\frac{e^{3u+v}}{3}+\frac{e^{6u-2v}}{6}\right)-f(v)\left(e^{3u+3v}-\frac{e^{3u+v}}{3}-\frac{e^{6u-2v}}{3}\right).\label{AA2}}
One notices that the choice $f(v)=0$ and $g(u)=2F_1e^{-6u}$ is consistent with both (\ref{AA1}) and (\ref{AA2}) as follows:
\eqn{a(u)=F_0,\hspace{10mm}b(v)=A=F_0+3F_1e^{-2v},}
where $F_0$ and $F_1$ are constants. Setting $F_1=0$ gives a Killing tensor proportional to the metric, whereas $F_1\neq0$ gives a nontrivial Killing tensor. $F_0=0$ and $F_1=1$ give the non-trivial rank-2 Killing tensor as
\eqn{K_{uu}=0,\hspace{5mm}K_{vv}=2\,e^{-6u}\,U^2,\hspace{5mm}K_{uv}=3\,e^{-2v}\,U.\label{K11}}


\subsubsection{Killing-Yano Tensors}
Since the minisuperspace has two dimensions, the maximum rank of Killing-Yano tensors is also two, and there exists at most only one rank-2 Killing-Yano tensor. The Killing-Yano equations $\nabla_{(\mu}f_{\nu)\rho}=0$ for $f_{uv}=-f_{vu}$ give
\eq{\nabla_uf_{uv}=\nabla_vf_{vu}=0.}
These equations give
\eqn{\partial_uf_{uv}-\Gamma^u_{uu}f_{uv}=0\Rightarrow f_{uv}=f(v)U,\\\partial_vf_{vu}-\Gamma^v_{vv}f_{vu}=0\Rightarrow f_{uv}=g(u)U.}
Therefore $f(v)=g(u)$ is a constant, and the associated Killing tensor $K_{\mu\nu}=f_{\mu\rho}f_{\nu}^{\rho}$ is just proportional to the metric:
\eq{K_{11}=K_{22}=0,\hspace{10mm}K_{12}=K_{21}\propto U.}


\subsection{Rank-4 Killing Tensors}
Similar to the case of rank-2, tedious calculations show that the following rank-4 symmetric tensor with constants $G_k$ is a Killing tensor
\eqn{K_{uuuu}&=&0,\label{K41}\\K_{uuuv}&=&81\,G_3\,e^{-2v}\,U^3\\K_{uuvv}&=&\Biggl(G_0+3G_1\,e^{-4v}+2G_2\,e^{-2v}\nonumber\\&&+G_3\left(2e^{6v}-12\,e^{4v}+18\,e^{2v}+54e^{6u-4v}-72e^{3u+v}\right)\Biggl)U^2,\\K_{uvvv}&=&\Biggl( 3G_1\,{{e}^{-6u-2v}}+G_2\, {{ e}^{-6u}}
\nonumber\\&&+G_3\left(18\,
{{e}^{-3u+v}}+27\,e^{-2v}-6\,{{ e}^{-3u+3v}}\right)\Biggr)U^3,\\K_{vvvv}&=&(2G_1\,e^{-12u}+12G_3\,e^{-6u})U^4.\label{K45}}
Note that if $G_0$ alone is nonzero, the Killing tensor is proportional to the symmetric product of two metrics; if $G_1$ alone is nonzero, the Killing tensor is proportional to the symmetric product of $K_{\mu\nu}$ from (\ref{K11}) with itself; if $G_2$ alone is nonzero, the Killing tensor is proportional to the symmetric product of $g_{\mu\nu}$ and $K_{\mu\nu}$; but if $G_3\neq0$, one gets a new nontrivial rank-4 Killing tensor. Here we shall set $G_0=G_1=G_2=0, G_3=16$. 


\section{Exact Solution}
As shown above, the two nontrivial invariants of motion are
\eqn{E_1&=&K_{\mu\nu}\frac{dx^{\mu}}{d\tau}\frac{dx^{\nu}}{d\tau}\nonumber\\&=&6e^{-2v}\,U\,\frac{du}{d\tau}\frac{dv}{d\tau}+2e^{-6u}\,U^2\,\left(\frac{dv}{d\tau}\right)^2,\label{E_1}}
\eqn{E_2&=&K_{\mu\nu\rho\sigma}\frac{dx^{\mu}}{d\tau}\frac{dx^{\nu}}{d\tau}\frac{dx^{\rho}}{d\tau}\frac{dx^{\sigma}}{d\tau}\nonumber\\&=&2^6\cdot3^4\,e^{-2v}\,U^3\left(\frac{du}{d\tau}\right)^3\frac{dv}{d\tau}+2^6\cdot3\,e^{-6u}\,U^4\,\left(\frac{dv}{d\tau}\right)^4\nonumber\\&&+2^6\cdot3\left(e^{6v}-6\,e^{4v}+9\,e^{2v}+27e^{6u-4v}-36e^{3u+v}\right)\,U^2\left(\frac{du}{d\tau}\right)^2\left(\frac{dv}{d\tau}\right)^2\nonumber\\&&+2^6\cdot3\left(6\,{{e}^{-3u+v}}+9\,e^{-2v}-2\,e^{-3u+3v}\right)\,U^3\,\frac{du}{d\tau}\left(\frac{dv}{d\tau}\right)^3,\label{E_2}}
where we are setting the Lagrange multiplier $\eta=1$ so that $\tau$ becomes the proper time along timelike geodesics in the minisuperspace metric, giving
\eq{\frac{4}{27}\,\frac{du}{d\tau}\frac{dv}{d\tau}U=1.\label{tau}}
This $\tau$ is not to be confused with the $\Lambda$-Taub-NUT time coordinate in the metric \eqref{Taub known}. By using (\ref{tau}), we can simplify the expressions of $E_1$ and $E_2$ to
\eq{E_1=\frac{81}{2}e^{-2v}+2\,e^{-6u}\,U^2\,\left(\frac{dv}{d\tau}\right)^2,\label{E_1 tau}}
\eqn{E_2&=&{3^{13}}\,e^{-2v}\,\left(\frac{dv}{d\tau}\right)^{-2}\nonumber\\&&+4\cdot{3^7}\left(e^{6v}-6\,e^{4v}+9\,e^{2v}+27e^{6u-4v}-36e^{3u+v}\right)\nonumber\\&&+2^4\cdot3^4\left(6\,{{e}^{-3u+v}}+9\,e^{-2v}-2\,e^{-3u+3v}\right)\,U^2\,\left(\frac{dv}{d\tau}\right)^2\nonumber\\&&+2^6\cdot3\,e^{-6u}\,U^4\,\left(\frac{dv}{d\tau}\right)^4.}
The right-hand side of (\ref{E_1 tau}) shows that $E_1>0$. The original constraint (\ref{constraint}) in null coordinates is
\eq{\frac{du}{d\zeta}\frac{dv}{d\zeta}=e^{-(3u+3v)}\,U.\label{original constraint}}
Thus by comparing this with (\ref{tau}), the relation between $d\zeta$ and $d\tau$ is
\eq{\frac{d}{d\zeta}=\frac{2U\,e^{-\frac{3}{2}(u+v)}}{3\sqrt{3}}\frac{d}{d\tau}.}
However one can see from the form of (\ref{E_1 tau}) that it becomes simplified if a new time coordinate $T$ is chosen as
\eq{\frac{d}{dT}=\frac{2}{9}\,e^{-3u+2v}\,U\,\frac{d}{d\tau}=\frac{1}{\sqrt{3}}\,e^{-\frac{3}{2}u+\frac{7}{2}v}\,\frac{d}{d\zeta}=2\sqrt{\Lambda}\,a(t)^3\,c(t)^{-1}\frac{d}{dt}.}
Then (\ref{E_1 tau}) can be rewritten as
\eqn{E_1&=&\frac{81}{2}e^{-2v}+\frac{81}{2}e^{-4v}\left(\frac{dv}{dT}\right)^2\nonumber\\&=&\frac{81}{2}e^{-2v}+\frac{81}{8}\left(\frac{d}{dT}e^{-2v}\right)^2,}
or
\eq{\frac{1}{4}\left(\frac{d}{dT}e^{-2v}\right)^2=\frac{2}{81}E_1-e^{-2v}.}
One can obtain the solution as
\eq{e^{-2v}=C^2-(T-T_0)^2\hspace{5mm}\left(C=\frac{\sqrt{2\,E_1}}{9}\right),\label{sol1}}
where the range of $T$ is
\eq{T_0-C\leq T\leq T_0+C.}
Note that $T_0$ is just a shift of time, so we choose $T_0=0$, and the inequalities above become equalities at past and future infinity for the proper time $t$ of the biaxial Bianchi IX spacetime metric. 

The constraint equation (\ref{tau}) then gives the solution for $u$. In terms of the $T$ coordinate, it becomes
\eqn{3\,e^{3u}\,e^{-7v}\,\frac{du}{dT}\frac{dv}{dT}&\equiv&-\frac{1}{2}\,e^{-5v}\,\left(\frac{d}{dT}e^{3u}\right)\left(\frac{d}{dT}e^{-2v}\right)\nonumber\\&=&1-e^{-2v}+e^{3u-5v}.}
By using (\ref{sol1}), one gets
\eq{\left(\frac{d}{dT}e^{3u}\right)=\frac{e^{3u}+e^{5v}-e^{3v}}{T},}
which has the solution
\eq{e^{3u}=B\,T+{\frac { (6\,{C}^{2}-8)T^4+(12\,C^2-9\,{C}^{4})\,T^2+3
\,{C}^{6}-3\,{C}^{4} }{ 3\,{C}^{6}\left( {C}^{2} -T^2\right) ^{\frac{3}{2}}}},\label{sol2}}
where $B$ is another constant which is related to $E_2$ by
\eq{E_2=4\cdot3^7\cdot\frac {9\,{B}^{2}{C}^{12}+36\,{C}^{4}-96\,{C}^{2}+64}{{C}^{6}}.}

Since $u$ and $v$ are given as explicit functions of $T$ by (\ref{sol1}) and (\ref{sol2}), the biaxial Bianchi IX metric (\ref{Bianchi metric}) can be written explicitly in terms of $T$ and the two parameters as
\eqn{ds^2&=&\frac{3}{\Lambda}\Biggl(-\frac{1}{3}\,e^{7v-3u}\,dT^2+\frac{1}{3}\,e^{2v}\,(d\theta^2+\sin^2\theta \;d\phi^2)+\frac{4}{3}\,e^{3u-v}\,(d\psi+\cos\theta\,d\phi)^2\Biggr)\nonumber\\&=&\frac{3}{\Lambda}\Biggl[-\frac{1}{3}(C^2-T^2)^{-\frac{7}{2}}\left(B\,T+{\frac { 6\,{T}^{4}{C}^{2}-9\,{T}^{2}{C}^{4}+3
\,{C}^{6}-8\,{T}^{4}+12\,{T}^{2}{C}^{2}-3\,{C}^{4} }{ 3\,{C}^{6}\left( {C}^{2} -T^2\right) ^{\frac{3}{2}}}}\right)^{-1}dT^2\nonumber\\&&+\frac{1}{3(C^2-T^2)}(d\theta^2+\sin^2\theta \;d\phi^2)\nonumber\\&&+\frac{4}{3}\sqrt{C^2-T^2}\left(B\,T+{\frac { 6\,{T}^{4}{C}^{2}-9\,{T}^{2}{C}^{4}+3
\,{C}^{6}-8\,{T}^{4}+12\,{T}^{2}{C}^{2}-3\,{C}^{4} }{ 3\,{C}^{6}\left( {C}^{2} -T^2\right) ^{\frac{3}{2}}}}\right)(d\psi+\cos\theta\,d\phi)^2\Biggr].\nonumber\\\label{Taub this}}
One can check that (\ref{Taub known}) and (\ref{Taub this}) coincide with each other by identifying their time coordinates and parameters as follows:
\eq{\frac{D_0}{4}(\tau^2+1)=\frac{1}{3}\,e^{2v}=\frac{1}{3(C^2-T^2)},}
\eq{D_0=\frac{4}{3C^2},\hspace{10mm}C_0=4BC^4.}


\section{Asymptotic Behaviour of the Bianchi IX model}

\subsection{Definition of Asymptotic States}

Let us consider the time evolution of a Bianchi IX spacetime without singularity, that is, a solution that is regular for all $-\infty<\zeta<\infty$. The motion in phase space is, in principle, determined by six initial conditions, namely $\al,\be,\ga$ and their conjugate momenta $\pi_{\al},\pi_{\be},\pi_{\ga}$. Since they are constrained by (\ref{constraint}), and since it is always possible to shift the time coordinate by a constant, one can get rid of two constants from the initial conditions. Hence the actual number of parameters is only four.

Consider those four parameters in the region where $\al\rightarrow\infty$. We assume that two of them are $\be$ and $\ga$, in other words, we assume that $\be$ and $\ga$ are asymptotically constants, and $\dot{\be},\dot{\ga}\rightarrow0$. Let us denote the other two asymptotic constants by $\cb,\cg$. As shall be shown shortly, $\pi_{\be}$ and $\pi_{\ga}$ diverge as $\al\rightarrow\infty$. In order to define $\cb$ and $\cg$, we first investigate the asymptotic behaviour of $\al$ and $\dot{\al}$, and then we calculate how $\dot{\be}$ and $\dot{\ga}$ damp as $\al$ grows. Since $\dot{\be},\dot{\ga}\rightarrow0$, (\ref{eqa}) and \eqref{constraint} give
\eqn{\ddot{\alpha}&=&3-3\dot{\al}^2-2Ve^{-2\al}\nonumber\\&=&1-\dot{\al}^2-2\dot{\be}^2-2\dot{\gamma}^2\nonumber\\&\rightarrow&1-\dot{\al}^2,}
where an overdot denotes the derivative with respect to the dimensionless time coordinate $\zeta$. The asymptotic solution is
\eq{\al=\ln(D\,e^{\zeta}+E\,e^{-\zeta}),\label{asy al}}
where $D$ and $E$ are constants that each depends on the choice of where $\zeta=0$, though their product $DE$ is invariant under constant shifts of $\zeta$ and is approximately the asymptotic value of $V/4$. In fact, a slightly better asymptotic form, with errors of the order of $e^{-5\alpha}$, can be shown to be
\eq{\alpha = \frac{1}{2}\ln{(F e^{2\zeta} + G + H e^{-2\zeta})},}
where $G$ is the asymptotic value of $V/2$ and where $FH$ is the
asymptotic value of $(1/16)V^2 - (1/8)[(\partial V/\partial\beta)^2 +
(\partial V/\partial\gamma)^2]$. Note that $D$ and $E$, or $F$, $G$, and $H$, at $\zeta\rightarrow\infty$ are in general different from $D$ and $E$, or $F$, $G$, and $H$, at $\zeta\rightarrow-\infty$. One can integrate (\ref{eqb}) to get
\eq{e^{3\al}\,\dot{\be}=\int d\zeta\left(\frac{1}{2}\frac{\partial V}{\partial \beta}e^{\alpha}\right).\label{db/dt 1}}
By assumption $\be$ and $\ga$ are asymptotically constants, so that one can approximate (\ref{db/dt 1}) by the following form:
\eq{e^{3\al}\,\dot{\be}\sim\frac{1}{2}\frac{\partial V}{\partial \beta}\int d\zeta\;e^{\alpha}.}
When (\ref{asy al}) is used, this is simplified to
\eq{\dot{\be}\sim\frac{1}{2}\frac{\partial V}{\partial \beta}\dot{\alpha}e^{-2\alpha}.\label{db2}}
This shows that for regular time evolution in which $\be$ and $\ga$ do not diverge as $\al$ goes to infinity, $\dot{\be}\rightarrow{\cal O}(e^{-2\al})$ for large $\al$, and similarly $\dot{\ga}\rightarrow{\cal O}(e^{-2\al})$. From the Lagrangian (\ref{ADM}), the conjugate momenta of $\be, \ga$ are
\eq{\pi_{\be}=-\,e^{3\al}\,\dot{\be},\hspace{5mm}\pi_{\ga}=-\,e^{3\al}\,\dot{\ga},}
which diverge as $\al$ grows. However there is a canonical transformation that gives parameters that are asymptotically constants. Let ($\tilde{\al},\tilde{\be},\tilde{\ga},\tilde{\pi}_{\al},\tilde{\pi}_{\be},\tilde{\pi}_{\ga}$) be a new canonical coordinates defined as
\eqn{\tilde{\al}&=&e^{2\al}-V,\\\tilde{\be}&=&\be\\\tilde{\ga}&=&\ga\\\tilde{\pi}_{\al}&=&\frac{\partial L}{\partial \dot{\tilde{\al}}}=\frac{1}{2}e^{\al}\dot{\al},\\\tilde{\pi}_{\beta}&=&\frac{\partial L}{\partial \dot{\tilde{\be}}}=-\,e^{3\alpha}\dot{\beta}+\frac{1}{2}\frac{\partial V}{\partial \beta}\dot{\alpha}e^{\alpha},\\\tilde{\pi}_{\ga}&=&\frac{\partial L}{\partial \dot{\tilde{\ga}}}=-\,e^{3\alpha}\dot{\ga}+\frac{1}{2}\frac{\partial V}{\partial \ga}\dot{\alpha}e^{\alpha},}
where we keep $\dot{\al}$ since the equation of motion (\ref{eqa}) is easier to evaluate than the analogous one for $\tilde{\al}$. It can then be shown that $\dot{\tilde{\pi}}_{\be}$ and $\dot{\tilde{\pi}}_{\ga}$ damp as ${\cal O}(e^{-\al})$. Indeed the time derivative of $\tilde{\pi}_{\be}$ is
\eq{\dot{\tilde{\pi}}_{\be}=-\frac{\partial V}{\partial \beta}(\dot{\beta}^2+\dot{\ga}^2)e^{\alpha}+\frac{1}{2}\frac{\partial^2 V}{\partial \beta^2}\dot{\beta}\dot{\alpha}e^{\alpha}+\frac{1}{2}\frac{\partial^2 V}{\partial \beta \partial \ga}\dot{\ga}\dot{\alpha}e^{\alpha}\sim{\cal O}(e^{-\al}),}
and similar for $\dot{\tilde{\pi}}_{\ga}$. Therefore if we choose $\cb=\tilde{\pi}_{\be}$ and $\cg=\tilde{\pi}_{\ga}$, all of the parameters ($\be,\ga,\cb,\cg$) are asymptotically constants. We call such a set of four asymptotic constants ($\be,\ga,\cb,\cg$) at $t\rightarrow-\infty$ an initial asymptotic state, while the set at $t\rightarrow+\infty$ is a final asymptotic state.


\subsection{Map from Past to Future Infinity}

Since asymptotic states are well-defined, one can consider a map $f$ from an initial asymptotic state to a final state,
\eq{f\;:\;(\be,\ga,\cb,\cg)_-\rightarrow\,(\be,\ga,\cb,\cg)_+.}
For the biaxial case, we know the exact solution, so that one can explicitly calculate $(\be,\cb)_{\pm}$ as
\eqn{\be&=&\frac{v-u}{2},\\\be_{\pm}&=&-\frac{1}{3}\ln\frac{4}{3\,C^2}\hspace{5mm}(T\rightarrow\pm C),}
\eqn{\cb&=&-\frac{1}{3}e^{3u-2v}\left(\frac{dv}{dT}-\frac{du}{dT}\right)-\frac{1}{3}(e^{3u-4v}-4e^{6u-7v})\left(\frac{dv}{dT}+\frac{du}{dT}\right),\\c_{\be\pm}&=&-\frac{B\,C^2}{3}\hspace{5mm}(T\rightarrow\pm C).}
Thus the map $f$ is the identity map in the biaxial case, which was suggested by the existence of two non-trivial Killing tensors in the minisuperspace. We perturbatively and numerically have found evidence that the map is no longer the identity in a triaxial model, but the exact form of the map has not yet been calculated.


\section*{Acknowledgment}
We would like to thank David Kubiznak  for helpful discussions. This research was supported by the Natural Sciences and Engineering Research Council of Canada.


\end{document}